\documentclass[twocolumn,pre]{revtex4}

\usepackage{amsfonts,amsmath,amssymb}
\usepackage{graphicx}

\begin{document}

\title{Noised induced phase transition in an oscillatory system with
dynamical traps}
\author{Ihor Lubashevsky}
\affiliation{Theory Department, General Physics Institute, Russian Academy of Sciences,
Vavilov Str. 38, Moscow, 119991 Russia}
\author{Morteza Hajimahmoodzadeh}
\affiliation{Faculty of Physics, M.~V.~Lomonosov Moscow State University, Moscow 119992,
Russia}
\author{Albert Katsnelson}
\affiliation{Faculty of Physics, M.~V.~Lomonosov Moscow State University, Moscow 119992,
Russia}
\author{Peter Wagner}
\affiliation{Institute of Transport Research, German Aerospace Center (DLR),
Rutherfordstrasse 2, 12489 Berlin, Germany.}
\date{\today }

\begin{abstract}
A new type of noised induced phase transitions is proposed. It occurs in noisy
systems with dynamical traps. Dynamical traps are regions in the phase space
where the regular ``forces'' are depressed substantially. By way of an example,
a simple oscillatory system $\{x,v=\dot{x}\}$ with additive white noise is
considered and its dynamics is analyzed numerically. The dynamical trap region
is assumed to be located near the $x$-axis where the ``velocity'' $v$ of the
system becomes sufficiently low. The meaning of this assumption is discussed.
The observed phase transition is caused by the asymmetry in the residence time
distribution in the vicinity of zero value ``velocity''. This asymmetry is due
to a cooperative effect of the random Langevin ``force'' in the trap region and
the regular ``force'' not changing the direction of action when crossing the
trap region.
\end{abstract}

\maketitle

\section{Introduction}

The ability of noise to produce order in systems, in particular, to induce
phase transitions is well established (see, e.g., Refs~\cite{In1,In2}). Such a
phase transition manifests itself in the phase-space density of the system
changing its structure, for example, the number of maxima. Noised-induced phase
transitions are distinguished from the classical ones by the fact that their
cause is not only the features of regular ``forces'' but also the action of
random Langevin ``forces''. As a result, in particular, the maxima of the
distribution function describing the noise-induced phases are not necessarily
related to the zero value of the regular ``forces''.

System of elements with motivated behavior, e.g., fish and bird swarms, car
ensembles on highways, stock markets, \textit{etc.} often display noise-induced
phase transitions. The formation of a new phase is caused by noise action (for
a review see Ref.~\cite{Hel1}). The theory of these phenomena is far from being
developed well.

The present Letter considers a certain class of such systems whose dynamics can
be described by two variables $x$, $v$ which perform a damped harmonic
oscillation near the equilibrium point $\{x=0,\,v=0\}$. However, the system
``cost'' of deviation from the equilibrium can differ substantially for these
variables. For example, in driving a car the control over the relative velocity
$v$ is of prime importance in comparison with the correction of the headway
distance $x$. So, under normal conditions a driver should eliminate the
relative velocity between her car and a car ahead first and only then correct
the headway. In markets the deviation from the supply-demand equilibrium
reflecting in price changes also has to exhibit faster time variations than,
e.g., the production cost determined by technology capabilities. In physical
systems this situation can be also met, e.g., in Pd-metal alloys charged with
hydrogen where the structure relaxation exhibits non-monotonic dynamics
\cite{K1,K2}. In these alloys hydrogen atoms and nonequilibrium vacancies form
long lived complexes affecting essentially the structure relaxation. Their
generation and disappearance governed, in turn, by the structure evolution
causes the non-monotonic dynamics which can be described in terms of dynamical
traps.

These observations lead to the concept of dynamical traps, a certain ``low''
dimensional region in the phase space where the main kinetic coefficients
specifying the characteristic time scales of the system dynamics become
sufficiently large in comparison with their values outside the trap region
\cite{we1,we2}. A trap region is not necessarily to be bounded in all the
dimensions and, in this case, it itself cannot lead to the formation of new
phases. Moreover, the equilibrium point can be absolutely stable, so without
noise there is only one phase in the system. The present Letter analyzes the
effect of noise on such a system and demonstrates that additive noise in a
system with dynamical traps is able to give rise to new phases. It should be
noted that in most models the phase transitions are caused by multiplicative
noise, however, additive noise also can give rise to these phenomena
\cite{WN0,WN1,WN2}.

\section{Noised oscillatory system with dynamical traps}

By way of example, the following system typically used to describe the
harmonic oscillatory dynamics is considered
\begin{eqnarray}
\frac{dx}{dt} &=&v\,,  \label{1.1a} \\
\frac{dv}{dt} &=&-\omega _{0}^{2}\Omega (v)\left[ x+\frac{\sigma }{\omega
_{0}}\,v\right] +\epsilon _{0}\xi _{v}(t)\,.  \label{1.1b}
\end{eqnarray}
Here $x$ and $v$ are the dynamical variables usually treated as a coordinate
and velocity of a certain particle, $\omega _{0}$ is the circular frequency of
oscillations provided the system is not affected by other factors, $\sigma$ is
the damping decrement, and the term $\epsilon _{0}\xi _{v}(t)$ in
equation~(\ref{1.1b}) is a random Langevin ``force'' of intensity $\epsilon
_{0}$ proportional to the white noise $\xi _{v}(t)$,
\begin{equation}
\left\langle \xi _{v}(t)\right\rangle =0\,,\quad \left\langle \xi _{v}(t)\xi
_{v}(t^{\prime })\right\rangle =\delta (t-t^{\prime })\,,  \label{1.2}
\end{equation}
with unit amplitude. The function $\Omega (v)$ describes the dynamical
trap effect arising in the vicinity of zero value velocity. For this
function, the following simple \textit{Ansatz} can be used
\begin{equation}
\Omega (v)=\frac{v^{2}+\triangle
^{2}\vartheta_{t}^{2}}{v^{2}+\vartheta_{t}^{2}}\,, \label{1.3}
\end{equation}
where the parameter $\vartheta_{t}$  characterizes the thickness of the trap
region and the parameter $\triangle \leq 1$ measures the trapping efficacy.
When $\triangle =1$ the dynamical trap effect is ignorable, for $\triangle =0$
it is most effective.

\begin{figure}
\begin{center}
  \includegraphics{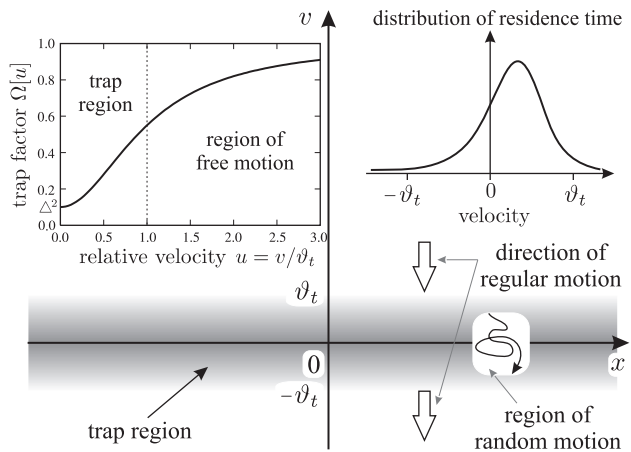}
\end{center}
\caption{Characteristic structure of the phase space $\{x,v\}$. The
  shadowed domain represents the trap region where the regular ``force''
  is depressed and the system motion is random. The regular ``force''
  depression is described by the factor $\Omega (v)$ illustrated in
  the left window. The essence of the trap effect on the system
  dynamics is shown in the right window. Outside the trap region the
  system dynamics is mainly regular.\label{Fig1}}
\end{figure}

The characteristic features of the given system are illustrated in
Fig.~\ref{Fig1}. The shadowed domain shows the trap region where the regular
``force'', the former term in Eq.~(\ref{1.1b}), is depressed. The latter is
described by the factor $\Omega (v)$ taking small values in the trap region
(for $\triangle \ll 1$). Inside the trap region the system is mainly governed
by the random Langevin ``force''. Outside the trap region it is approximately
harmonic.

In order to analyze the system dynamics a dimensionless time $t$ and the
dynamical variables $\eta $ and $u$ are used. Namely, the time $t$ is measured
in units of $1/\omega _{0}$, i.e., $t\rightarrow t/\omega _{0}$ and the units
of the coordinate $x$ and the velocity $v$ are $\vartheta_{t}/\omega _{0}$ and
$\vartheta_{t}$, respectively. So, by introducing the new variables
\begin{equation*}
\eta =\frac{x\omega _{0}}{\vartheta_{t}}\quad \text{and}\quad
u=\frac{v}{\vartheta_{t}}
\end{equation*}
the dynamical equations~(\ref{1.1a}), (\ref{1.1b}) read (for the dimensionless
time $t$)
\begin{equation}
\frac{d\eta }{dt}=u\,,\quad \frac{du}{dt}=-\Omega \lbrack u]\left( \eta +\sigma
u\right) +\epsilon \xi (t)\,,  \label{1.4}
\end{equation}
where the noise $\xi (t)$ obeys conditions like equalities~(\ref{1.2}), the
parameter $\epsilon =\epsilon _{0}/(\sqrt{\omega _{0}}\vartheta_{t})$, and the
function
\begin{equation*}
\Omega \lbrack u]=\frac{u^{2}+\triangle ^{2}}{u^{2}+1}\,.
\end{equation*}

Without noise, this system has only one stationary point $\{\eta =0,u=0\}$
being stable because it possesses a Liapunov function
\begin{equation}
\mathcal{H}(\eta ,u)=\frac{\eta ^{2}}{2}+\frac{u^{2}}{2}+\frac{1-\triangle
^{2}}{2}\ln \left( \frac{u^{2}+\triangle ^{2}}{\triangle ^{2}}\right) .
\label{1.5}
\end{equation}
This Liapunov function attains the absolute minimum at the point
$\{\eta =0,u=0\}$ and obeys the inequality
\begin{equation}
\frac{d\mathcal{H}(\eta ,u)}{dt}=-\sigma u^{2}<0\quad \text{for}\quad u\neq
0\,.  \label{1.6}
\end{equation}
In particular, if $\sigma =0$ and $\epsilon =0$ then
function~(\ref{1.5}) is the first integral of the system. In what
follows, the values $\sigma $ and $\epsilon $ will be treated as small
parameters.

The present Letter demonstrates the fact that the noise $\xi (t)$ can cause a
phase transition in the given system. It manifests itself in that the
distribution function $\mathcal{P}(\eta ,u)$ changes form from a unimodal to a
bimodal one. The dynamics of system~(\ref{1.4}) was analyzed numerically using
a high order stochastic Runge-Kutta method \cite{RK1} (see also
Ref.~\cite{RK2}). The distribution function $\mathcal{P}(\eta ,u)$ was
calculated numerically by finding the cumulative time during which the system
is located inside a given mesh on the $\left( \eta ,u\right) $-plane for a path
of a sufficiently long time of motion, $t\approx 500000$. The size of mesh was
chosen to be about 1\% of the dimension characterizing the system location on
the $\left( \eta ,u\right) $-plane.

\begin{figure}
\begin{center}
  \includegraphics{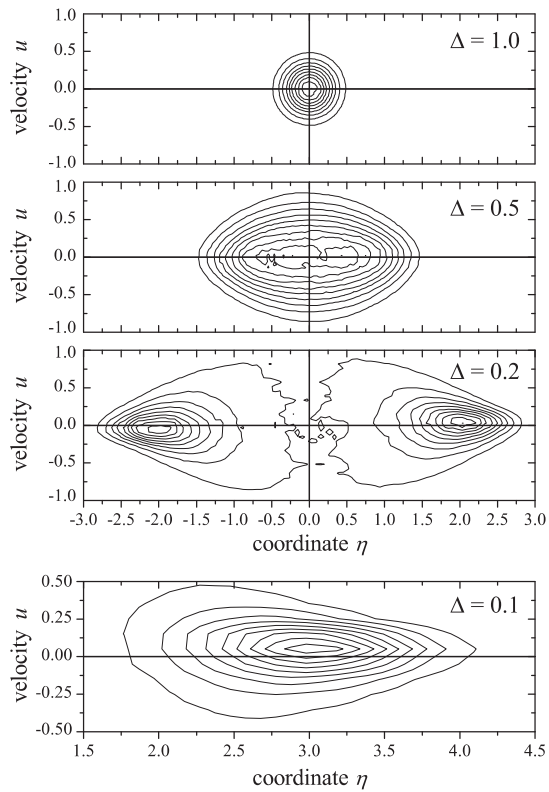}
\end{center}
\caption{Evolution of the distribution function
   $\mathcal{P}(\protect\eta ,u)$ (shown by level contours) as the
   parameter $\triangle$ decreases. In numerical calculations the
   values $\sigma =0.1$ and $\epsilon =0.1$ were used. The lower
   window depicts only one maximum of the distribution
   function.\label{Fig2}}
\end{figure}

\begin{figure}
\begin{center}
  \includegraphics{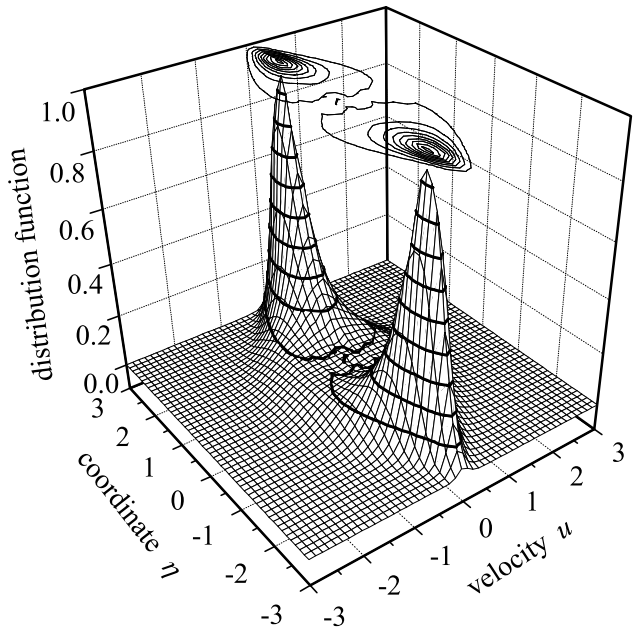}
\end{center}
\caption{The form of the distribution function
   $\mathcal{P}(\protect\eta ,u)$ for the parameters $\sigma =0.1$,
   $\epsilon =0.1$, and $\triangle =0.2$.\label{Fig3}}
\end{figure}

The evolution of the distribution function $\mathcal{P}(\eta ,u)$ is
shown in Fig.~\ref{Fig2} in the form of the level contours dividing
the variation scale into ten equal parts. The upper window corresponds
to the case of $\triangle =1$ where the trap effect is absent and the
distribution function is unimodal. The third window illustrates the
case when the distribution function has the well pronounced bimodal
shape shown also in Fig.~\ref{Fig3}.  Comparing the three upper
windows in Fig.~\ref{Fig2} it becomes evident that there is a certain
relation $\Phi _{c}(\triangle ,\sigma ,\epsilon )=0$ between the
parameters $\triangle $, $\sigma $, and $\epsilon $ when the system
undergoes a second order phase transition which manifests itself in
the change of the shape of the phase space density $\mathcal{P}(\eta
,u)$ from unimodal to bimodal. In particular, for $\sigma =0.1$ and
$\epsilon =0.1$ the critical value of the parameter $\triangle $ is
$\triangle _{c}(\sigma ,\epsilon )\approx 0.5$ as seen in the second
window.

\section{Mechanism of the phase transition}

\begin{figure}
\begin{center}
\includegraphics{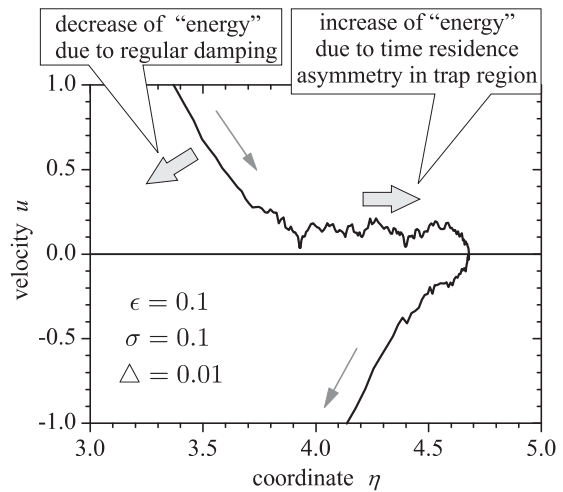}
\end{center}
\caption{A typical fragment of the system path going through the
   trap region. The parameters $\sigma =0.1$, $\epsilon =0.1$, and
   $\triangle =0.01$ were used in numerical simulations in order to
   make the trap effect more pronounced.}  \label{Fig4}
\end{figure}

To understand the mechanism of the noised induced phase transition observed
numerically in the given system, consider a typical fragment of the system
motion through the trap region for $\triangle \ll 1$ that is shown in
Fig.~\ref{Fig4}. When it goes into the trap region $\mathcal{Q} _{t}$,
$-\vartheta_{t}\ll v\ll \vartheta_{t}$, the regular ``force'' $\Omega \lbrack
u]\left(\eta +\sigma u\right) $ containing the trap factor $\Omega \lbrack u]$
and governing the regular motion becomes small. So inside this region the
system dynamics becomes random due to the remaining weak Langevin ``force''
$\epsilon \xi (t)$. However, the boundaries $\partial_{+}\mathcal{Q}_{t}$
(where $v\sim \vartheta_{t}$) and $\partial _{-}\mathcal{Q}_{t}$ (where $v\sim
-\vartheta_{t}$) are not identical in properties with respect to the system
motion. At the boundary $\partial _{+}\mathcal{Q}_{t}$ the regular ``force''
leads the system inwards the trap region $\mathcal{Q}_{t}$, whereas at the
boundary $\partial _{-}\mathcal{Q}_{t}$ it causes the system to leave the
region $\mathcal{Q}_{t}$. Outside the trap region $\mathcal{Q}_{t}$ the regular
``force'' is dominant. Thereby, from the standpoint of the system motion inside
the region $\mathcal{Q}_{t}$, the boundary $\partial _{+}\mathcal{Q}_{t}$ is
``reflecting'' whereas the boundary $\partial _{-}\mathcal{Q} _{t}$ is
``absorbing''.

As a result the distribution of the residence time at different points of the
region $\mathcal{Q}_{t}$ should be asymmetric, as schematically shown in
Fig.~\ref{Fig1}(the right window). This asymmetry is also seen in the
distribution function $\mathcal{P}(\eta ,u)$ obtained numerically. Its maxima
are located at the points with non-zero values of the velocity, which is
clearly visible in the lower window of Fig.~\ref{Fig2}. Therefore, during
location inside the trap region the mean velocity of the system must be
positive and it tends to go away from the origin. This effect gives rise to an
increase in the ``energy'' $\mathcal{H}(\eta ,u)$. Outside the trap region the
``energy'' $\mathcal{H}(\eta ,u)$ decreases according to
expression~(\ref{1.6}). So, when the former effect becomes sufficiently strong,
i.e., the random ``force'' intensity $\epsilon $ exceeds a certain critical
value, $\epsilon >\epsilon _{c}(\triangle,\sigma )$, the distribution function
$\mathcal{P}(\eta ,u)$ becomes bimodal.

The system location with respect to the velocity $v$ is due to the regular
``force'' being sufficiently strong outside the trap region, so the system
spends the main time inside this region. Its location with respect to the
coordinate $x$ is caused by the fact that the region where the Langevin
``force'' mainly affects the system dynamics decreases in thickness as the
coordinate $x$ increases. The latter tendency takes place because the regular
``force'', the first term in Eq.~(\ref{1.1b}), is proportional to $x$. In fact,
the thickness $U(\eta)$ of the trap region in the vicinity of the point
$\{\eta,u=0\}$ can be estimated using the condition of the equality of the
characteristic times, $\tilde{t}_s$ and $\tilde{t}_d$, during which the system
crosses the trap region under action of the regular ``force'' and the random
Langevin ``force''. So
\begin{equation*}
\tilde{t}_s\sim \frac{U}{\Omega[U]\eta}\sim \tilde{t}_d\sim
\frac{U^2}{\epsilon^2}\,.
\end{equation*}
and setting for the sake of simplicity $\triangle=0$ we get the estimate
\begin{equation*}
 U(\eta)\sim \epsilon^{2/3}\,\eta^{-1/3}\,.
\end{equation*}
Moreover, let the mean velocity in the trap region caused by the residence time
asymmetry be about $U(\eta)$. Then the characteristic increase $\delta\eta$ of
the coordinate $\eta$ got by the system when crossing the trap region is
estimated as $\delta\eta\sim U\tilde{t}_s\sim 1/\eta$. Thereby the dynamical
trap effect becomes weaker as the ``energy'' $\mathcal{H}(\eta ,u)$ increases.
By contrast, according to Exp.~(\ref{1.6}) the higher the ``energy'', the
stronger its dissipation caused by the regular ``force''.

\section{Conclusion}

The present paper has considered noise-induced phase transitions in systems
with dynamical traps. By way of example, a simple oscillatory system
$\{x,v=\dot{x}\}$ is studied when the trap region is located in the vicinity of
the $x$-axis and without noise the stationary point $\{x=0,v=0\}$ is absolutely
stable. For this system as shown numerically the additive white noise can cause
the phase-space density to take the bimodal shape.

In contrast to the classical phase transitions the position of the new
noise-induced phases is not specified by the zero-values of regular ``forces''
even approximately. The cause of the observed phase transition is the asymmetry
of the residence time distribution inside the trap region. This asymmetry is
due the cooperative effect of the regular ``force'' outside the trap region and
the random Langevin ``force'' inside it. The regular ``force'' does not change
the direction when crossing the trap region, inside this region it is depressed
only. As a result, for the motion inside the trap region one of its boundaries
is ``reflecting'', whereas the other is ``absorbing'', which induces the
residence time asymmetry. The latter gives rise to increase in the system
``energy''. Outside the trap region the regular ``force'' causes the ``energy''
to decrease.

\acknowledgments

These investigations were supported in part by RFBR~Grants 01-01-00389,
02-02-16537, INTAS Grant~00-0847, and Russian Program ``Integration'',
Project~B0056.

\end{document}